\documentclass[letterpaper,10pt]{article}
\usepackage{osameet2}
\usepackage[amssymb]{SIunits}
\usepackage[cmex10]{amsmath} 
\usepackage[latin1]{inputenc}
\usepackage{xcolor}
\usepackage{amsmath, amssymb,amsthm}
\usepackage{bm}
\usepackage{enumitem}

\usepackage{graphicx}
\usepackage[caption=false, font=footnotesize]{subfig}
\usepackage{xspace} 

\usepackage{multirow}
\usepackage{multicol}
\usepackage{floatrow}
\usepackage{wrapfig}
\newfloatcommand{capbtabbox}{table}[][\FBwidth]

\usepackage{titlesec}
\titlespacing\subsection{0pt}{4pt plus 2pt minus 2pt}{3pt plus 2pt minus 2pt}

\newcommand{\D}{\mathrm{d}}
\newcommand{\define}{\triangleq}
\newcommand{\vect}[1]{\boldsymbol{#1}}

\newcommand{\imag}{\ensuremath{\jmath}}

\newcommand{\StepSize}{\ensuremath{\delta}}
\newcommand{\NumSteps}{\ensuremath{M}}

\newcommand{\nlop}[1]{\ensuremath{\bm{\sigma}_{#1}}}
\newcommand{\DBP}{\ensuremath{\text{DBP}}}

\newcommand{\FFTm}{\mat{W}}

\DeclareMathOperator{\diag}{\text{diag}}

\newcommand{\Rsymb}{\ensuremath{f_\text{symb}}}
\newcommand{\Rsamp}{\ensuremath{f_\text{samp}}}

\renewcommand{\imag}{\ensuremath{j}}
\renewcommand{\vect}[1]{\ensuremath{\mathbf{#1}}}
\newcommand{\mat}[1]{\ensuremath{\mathbf{#1}}}

\newcommand{\transpose}{\intercal}

\newcommand{\Lsp}{\ensuremath{L_{\text{sp}}}}
\newcommand{\Nsp}{\ensuremath{N_\text{sp}}}

\newcommand{\norm}[1]{\|#1\|}

\begin{document}

\title{\LARGE Nonlinear Interference Mitigation\\ via Deep
Neural Networks}

\newcommand{\aff}[1]{\textsuperscript{(#1)}}

\vspace{-0.4cm}

\author{Christian H\"ager\aff{1,2} and
	Henry D.~Pfister\aff{2}
}
\address{\textsuperscript{(1)}Department of Electrical Engineering, Chalmers
	University of Technology, SE-41296 G\"oteborg, Sweden, \\
	\textsuperscript{(2)}Department of Electrical and Computer Engineering, Duke
	University, Durham, NC, 27708, US
	}
\email{(e-mail: christian.haeger@chalmers.se, henry.pfister@duke.edu)}

\vspace{-0.2cm}

%
\begin{abstract}
	A neural-network-based approach is presented to efficiently
	implement digital backpropagation (DBP). For a
	32$\,\times$100$\,\,$km fiber-optic link, the resulting
	``learned'' DBP significantly reduces the complexity compared to
	conventional DBP implementations. 
\end{abstract}


\ocis{(060.0060) Fiber optics and optical communications, (060.2330)
Fiber optics communications.}

\vspace*{-0.25cm}

\section{Introduction}


Nonlinear interference (NLI) is a significant challenge in high-speed
fiber-optic communication systems. One approach to mitigate NLI is by
solving the nonlinear Schr\"odinger equation (NLSE) with negated fiber
parameters as part of the receiver processing. This is commonly
referred to as digital backpropagation (DBP) \cite{Ip2008}. Several
authors have highlighted the large computational burden associated
with DBP and proposed various techniques to reduce its complexity
\cite{Du2010, Tao2011, Li2011, Shen2011, Jarajreh2015,
Giacoumidis2015}. In essence, the task is to approximate the solution
of a partial differential equation using as few computational
resources as possible. We approach this problem from a
machine-learning perspective. Compared to previous work in
\cite{Shen2011, Jarajreh2015, Giacoumidis2015}, we focus on deep
neural networks (NNs), which have attracted tremendous interest in
recent years \cite{LeCun2015}. 

One issue with standard deep NNs is the absence of clear guidelines
for the network design, e.g., choosing the number of network layers.
This issue can be addressed by using an existing algorithm for the
considered task and interpreting its associated computation graph as a
blueprint for the NN. Since many algorithms are iterative, this
procedure often entails ``unrolling'' the iterations which then form
the layers in the ensuing network. This approach has been proposed for
sparse signal recovery \cite{Gregor2010} and also applied in other
areas, e.g., decoding linear codes via belief propagation
\cite{Nachmani2016a}. In this paper, we show that similar ideas can be
applied to the NLI mitigation problem.  In particular, we exploit the
fact that the unrolled split-step Fourier method (SSFM) has
essentially the same functional form as a deep NN. 

\section{System model}
\label{sec:system_model}

The data is pulse modulated to get $x(t) = \sum_{k=1}^m x_{k}
p(t-k/\Rsymb)$, where $x_1, \dots, x_m \in \mathcal{X}$ are symbols
from a complex signal constellation $\mathcal{X}$, $p(t)$ is the pulse
shape, and $\Rsymb$ is the symbol rate. The signal $x(t)$ is launched
into an optical fiber and propagates according to the NLSE
$\frac{\partial u(t,z)}{\partial z} = (- \frac{\alpha}{2} - \imag
\frac{\beta_2}{2} \frac{\partial^2 }{\partial t^2}) u(t,z)+ \imag
\gamma |u(t,z)|^2 u(t,z)$, where $u(t,0) = x(t)$ and $\alpha, \beta_2,
\gamma$ are the attenuation, dispersion, and nonlinearity parameters,
respectively.  After distance $z=L$, the signal $u(t,L)$ is low-pass
(LP) filtered and sampled at $t = k / \Rsamp$ to give the observation
vector $\vect{y} = (y_1, \dots, y_n)^\transpose \in \mathbb{C}^n$. 


\subsection{Digital backpropagation}

In the absence of noise, the symbol vector $\vect{x} \define (x_1, \dots,
x_m)^\transpose $ can be recovered by solving the NLSE with negated
fiber parameters $\alpha,\beta_2, \gamma$, followed by a digital
matched filter (MF). In the following, we use the time-discretized
NLSE \vspace{-0.2cm}
\begin{align}
	\label{eq:discretized_nlse}
	\frac{\D \vect{u}(z)}{\D z} 
	= \mat{A} \vect{u}(z) -
	\imag \gamma |\vect{u}(z)|^2 \circ \vect{u}(z), 
\end{align}\\[-0.4cm]
where $\vect{A} = \FFTm^{-1} \diag(H_1, \dots, H_n) \FFTm$, $\FFTm$ is
the $n\times n$ discrete Fourier transform (DFT) matrix, $H_k =
\frac{\alpha}{2} + \imag \frac{\beta_2}{2} (2 \pi f_k)^2$, $f_k$ is
the $k$-th DFT frequency, and $\circ$ denotes element-wise vector
multiplication. Consider now the initial value problem defined by
\eqref{eq:discretized_nlse} and $\vect{u}_0 \define \vect{u}(0) =
\vect{y}$. In particular, let the mapping $\DBP : \mathbb{C}^n \to
\mathbb{C}^n$ be such that $\vect{u}(L) = \DBP(\vect{y})$. Our goal is
to implement this mapping in a computationally efficient way. 

\subsection{Split-step Fourier method}

A popular numerical method to implement DBP is the SSFM. Note that for
$\gamma = 0$, \eqref{eq:discretized_nlse} is linear with $\vect{u}(z)
= \mat{A}_z \vect{u}_0$, where $\mat{A}_z \define e^{z \mat{A}} =
\FFTm^{-1} \diag(e^{z H_1}, \dots, e^{z H_n}) \FFTm$. Moreover, for
$\beta_2 = 0$ and $\alpha=0$, the solution is $\vect{u}(z) =
\nlop{z}(\vect{u}_0)$, where $\nlop{z} : \mathbb{C}^n \to
\mathbb{C}^n$ is defined as the element-wise application of
$\sigma_z(x) = x e^{-\imag \gamma z |x|^2 }$. After conceptually
dividing the fiber into $\NumSteps$ segments of length $\StepSize = L
/ \NumSteps$, the (symmetric) SSFM is defined by $\vect{u}_{k} =
\mat{A}_{\StepSize/2} \nlop{\StepSize}\left( \mat{A}_{\StepSize/2} \vect{u}_{k-1}\right)$ for
$k = 1, \dots, \NumSteps$, where $\vect{u}_M$ serves as an estimate
$\vect{u}_{M} \approx \vect{u}(M \StepSize) = \text{DBP}(\vect{y})$
for the backpropagated signal vector. Unrolling all iterations gives
\vspace{-0.1cm}
\begin{align}
	\label{eq:ssfm}
	\vect{u}_{\NumSteps} =
	\mat{A}_{\StepSize/2}
	\nlop{\delta}( \vect{A}_\StepSize \dots \nlop{\delta}
	(\vect{A}_\StepSize
	\nlop{\delta}(\vect{A}_{\StepSize/2}	
	\vect{u}_0))),
\end{align}
where we used $\mat{A}_{\StepSize/2}\mat{A}_{\StepSize/2} =
\mat{A}_{\StepSize}$. The accuracy of \eqref{eq:ssfm} can be increased
by decreasing the step size $\StepSize$. However, this also increases
the complexity by increasing the number of steps $M$, which may be
prohibitive for practical implementations. 

\section{Deep neural networks}


Deep (feed-forward) NNs map an input vector $\vect{a}$ to an output
vector $\vect{b} = \bm{\rho}^{(\ell)} (\vect{B}^{(\ell)} (\dots
{\bm{\rho}^{(1)}} ( \vect{B}^{(1)}(\vect{a}))))$, where $\ell$ is the
number of layers, $\vect{B}^{(1)}, \dots, \vect{B}^{(\ell)}$ are
linear (or affine) functions, and $\bm{\rho}^{(1)}, \dots,
\bm{\rho}^{(\ell)}$ are nonlinear functions \cite{LeCun2015}. The
linear functions are given by $\vect{B}^{(k)}(\vect{c}) =
\vect{W}^{(k)} \vect{c} + \vect{b}^{(k)}$ for all $k$, where the
matrices $\vect{W}^{(1)}, \dots, \vect{W}^{(\ell)}$ and vectors
$\vect{b}^{(1)},\dots,\vect{b}^{(\ell)}$ are referred to as the
network weights and biases, respectively. The nonlinear functions
typically correspond to the element-wise application of some smooth
(i.e., differentiable) function $\rho(x)$, e.g., the logistic or
sigmoid function. 

For NNs, all involved quantities (i.e., the network input, output,
weights, and biases) are typically real-valued rather than
complex-valued. Moreover, the dimension of the weight matrices and
bias vectors may vary across layers. Besides these differences, deep
NNs and the SSFM in \eqref{eq:ssfm} have essentially the same
functional form: in both cases one alternates between the application
of linear operators and simple element-wise nonlinear operators. 

Deep NNs have achieved record-breaking performance for various machine
learning tasks such as speech or object recognition \cite{LeCun2015}.
Such tasks are seemingly unrelated to nonlinear signal propagation and
one may wonder if the similarity to the SSFM is merely a coincidence.
In that regard, some authors argue that deep NNs perform well because
their functional form matches the hierarchical or Markovian structure
that is present in most real-world data \cite{Lin2017}. Indeed, the
SSFM can be seen as an example where such a structure arises, i.e., by
decomposing the physical process described by
\eqref{eq:discretized_nlse} into a hierarchy of elementary steps.  

\section{Learned digital backpropagation}




\begin{figure}
	\centering \includegraphics{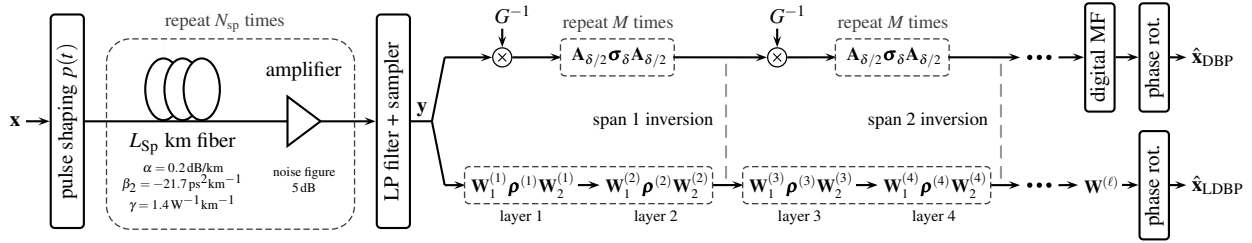} \caption{Block diagram
	showing the end-to-end system model (LP: low-pass, MF: matched
	filter). The top processing branch corresponds to DBP via the SSFM
	and the bottom branch to the proposed learned DBP (LDBP), obtained
	by ``unrolling'' the SSFM. Prior to the parameter optimization via deep
	learning, LDBP has the same performance as DBP assuming the same
	number of steps/layers per span. \vspace*{-0.2cm} }
	\label{fig:block_diagram}
	\vspace{-0.2cm}
\end{figure}

\newcommand{\xhatdbp}{\hat{\vect{x}}_{\text{DBP}}}
\newcommand{\xhatldbp}{\hat{\vect{x}}_{\text{LDBP}}}

The main idea in this paper is to interpret the SSFM in
\eqref{eq:ssfm} as a blueprint for a complex-valued deep NN and
optimize the network parameters using machine learning tools. We refer
to the resulting method as learned DBP (LDBP). 

In the following, we consider a multi-span system where the optical
link consists of $\Nsp$ spans of length $\Lsp$, as shown in
Fig.~\ref{fig:block_diagram}. An optical amplifier is inserted after
each span to compensate for the signal attenuation. In the SSFM, this
is accounted for by including multiplicative factors $G^{-1} =
e^{-\frac{\alpha}{2} \Lsp}$ as shown in the top processing branch in
Fig.~\ref{fig:block_diagram}. The estimated symbol vector $\xhatdbp
\in \mathbb{C}^m$ is obtained by applying a digital MF followed by a
phase-offset rotation.  



\subsection{Neural network parameters}


The bottom branch in Fig.~\ref{fig:block_diagram} shows an example for
the NN in LDBP based on unrolling the SSFM with $M = 2$ steps per span
(StPS). Each network layer comprises two weight matrices
$\smash{\mat{W}^{(i)}_1, \mat{W}^{(i)}_2 \in \mathbb{C}^{n \times n}}$
(no bias vectors are used). For the $i$-th layer, the nonlinearity
$\smash{\bm{\rho}^{(i)} : \mathbb{C}^n \to \mathbb{C}^n}$ acts
element-wise using $\rho^{(i)}(x) = x e^{-\imag \alpha_i |x|^2}$ in
each dimension, where $\alpha_i,x \in \mathbb{C}$. The function
$\rho^{(i)}$ is differentiable and can thus be used with standard
gradient-based optimization algorithms for deep learning. The MF is
accounted for by inserting an additional linear layer with weight
matrix $\mat{W}^{(\ell)} \in \mathbb{C}^{n\times m}$, where $\ell = M
\Nsp + 1$ is the total number of network layers. 

The network parameters are $\theta = \{\smash{\vect{W}^{(1)}_1, \dots,
\vect{W}^{(\ell-1)}_1, \vect{W}^{(1)}_2, \dots, \vect{W}^{(\ell-1)}_2,
\vect{W}^{(\ell)}}, \alpha_1, \dots, \alpha_{\ell-1}\}$. For the
optimization, all weight matrices are restricted to an equivalent
circular convolution with a symmetric filter of length $2K+1$. That
is, the matrix rows are circularly shifted versions of $(h_{-K},
\dots, h_{-1}, h_0, h_{1}, \dots, h_{K}, 0, \dots, 0)$, where $h_i \in
\mathbb{C}$ and $h_{-i} = h_{i}$ for $i = 1, 2, \dots, K$. This
reduces the number of free (complex-valued) parameters per weight
matrix from $n^2$ to $K+1$, where $K \ll n$. This restriction also
implies that LDBP is fully compatible with a potential time-domain
filter implementation \cite{Fougstedt2017}. The weight matrices are
initialized by using the appropriately zeroed versions of
$\smash{\mat{A}_{\StepSize/2}}$, multiplied by $G^{-1}$ for
$\smash{\mat{W}_1^{(i)}}$ in the first layer of each span.
Furthermore, we initialize $\alpha_i = \gamma \delta$ and set
$\smash{\mat{W}^{(\ell)}}$ to the MF. Thus, prior to the parameter
optimization, LDBP has the same performance as DBP with the same
number of StPS (assuming sufficiently large $K$).

\subsection{Objective function and optimization procedure}

NNs are trained by using many pairs of input and desired--output
examples and adjusting the network parameters $\theta$ such that some
predefined loss function between the network output and the desired
output decreases. For LDBP, the network input is $\vect{y} \in
\mathbb{C}^n$ and the desired output is the transmitted symbol vector
$\vect{x} \in \mathcal{X}^m$. As a loss function, we use the mean
squared error $\norm{\vect{x}-\xhatldbp}$, where $\norm{\vect{x}}
\define \sum_{i=1}^{m} |x_i|^2$ and $\xhatldbp \in \mathbb{C}^m$
denotes the NN output of LDBP. Assuming that $\|\vect{x}\|$ is
constant for all $\vect{x}$, this is equivalent to maximizing the Q-factor
$\norm{\vect{x}} / \norm{\vect{x} - \xhatldbp}$. We used TensorFlow to
implement the NN and optimize the network parameters. For the
optimization, we use the built-in \emph{Adam} optimizer,
which performs stochastic gradient descent with 30
input--output pairs $(\vect{y}, \vect{x})$ (i.e., mini-batches) per
optimization step.  

\begin{wrapfigure}[15]{r}{8.0cm}
	\vspace{-0.45cm}
	\centering
	\includegraphics{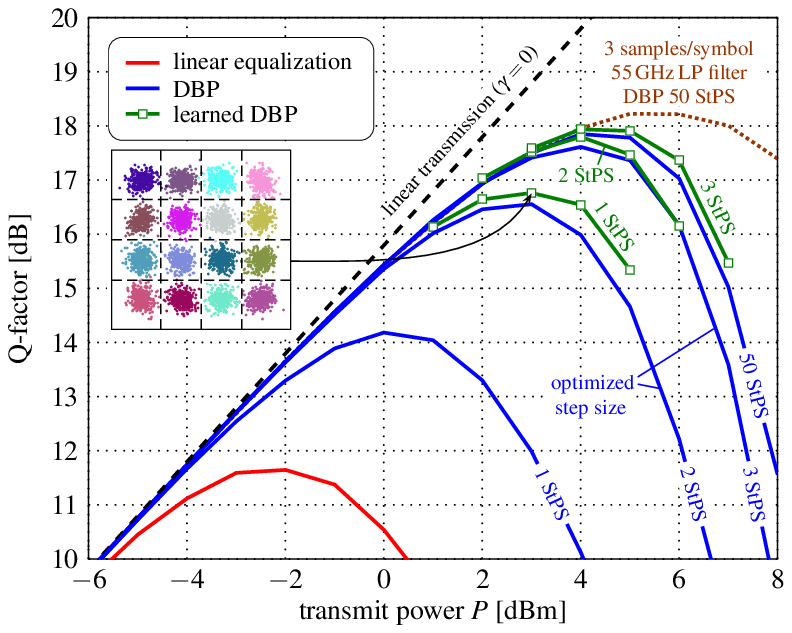}
	\vspace*{-0.20cm}
	\caption{Results}
	\label{fig:Qresults}
\end{wrapfigure}

\noindent Our system setup is such that the training data is
generated on-the-fly utilizing all CPU cores, whereas the
gradient-descent optimization is performed in parallel on the same
machine using an NVIDIA Tesla K40c GPU. 

\section{Numerical results}

We assume $16$-QAM transmission at $20\,$Gbaud using root-raised
cosine pulses (roll-off factor 0.1). For the optical link, we set
$\Nsp = 32$ and $\Lsp = 100\,$km with fiber parameters shown in
Fig.~\ref{fig:block_diagram}. Brick-wall LP filtering with $35\,$GHz
bandwidth is applied before sampling at $f_\text{samp} = 40\,$GHz,
i.e., 2 samples/symbol are used for the receiver processing.  Forward
propagation is simulated using 6 samples/symbol and $50$ StPS in the
SSFM (increasing either value did not affect the results).  We
consider LDBP with 1, 2, and 3 StPS, where the filter memory $K$ is
set to $12$, $8$, and $6$, respectively. The resulting Q-factor is
shown in Fig.~\ref{fig:Qresults}. During the optimization, the
transmit power for a particular input--output pair is chosen randomly
from the set of powers that is shown by the markers in
Fig.~\ref{fig:Qresults}, e.g., $P \in \{1,2,3,4,5\}$ dBm for LDBP with
1 StPS. As a comparison, we show the performance of DBP via the SSFM
with 1, 2, 3, and 50 StPS.  For 2 and 3 StPS, we used an optimized
(non-uniform) step size per span. At the optimal transmit power, LDBP
with 1 StPS provides a Q-factor of $16.8\,$dB compared to $16.6\,$dB
for DBP with 2 StPS. Taking the number of StPS as a rough measure of
complexity, LDBP thus gives a 50\% complexity reduction over DBP for
comparable performance. Moreover, due to the relatively short filter
memory $K$, a time-domain implementation may be preferred over the use
of Fourier transforms to implement the linear convolutions
\cite{Fougstedt2017}. As seen in Fig.~\ref{fig:Qresults}, using more
StPS gives diminishing returns in terms of the achievable Q-factor for
both LDBP and DBP. However, it is noteworthy that LDBP with 3 StPS
slightly outperforms DBP with 50 StPS in the nonlinear regime
(although the optimal Q-factor is essentially the same). This is
likely due to the fact that a significant part of the broadened
spectrum of the received waveform is filtered out before processing.
While LDBP compensates somewhat for this effect, the missing spectrum
is not backpropagated correctly when using DBP. In that case, a larger
receiver bandwidth is necessary as shown in Fig.~\ref{fig:Qresults} by
the dotted line. 


\section{Conclusion and future work}

We have proposed an approach for NLI mitigation using deep NNs,
where the network design is based on unrolling the SSFM. Compared to
standard ``black-box'' deep NNs, this approach leads to clear
hyperparameter choices (e.g., number of network layers, type of
nonlinearity, etc.) and also provides a good initialization for the
gradient-based optimization. The resulting learned DBP significantly
reduces the complexity compared to conventional DBP implementations. 

One of the most appealing features of NNs is that they can adapt to
real-world imperfections that are not included in analytical models
such as \eqref{eq:discretized_nlse}. For future work, it would
therefore be interesting to perform the parameter optimization based
on experimental data. It could also be viable to explore NN designs
for LDBP that include modified nonlinear functions, e.g., with
additional filtering steps \cite{Du2010}, in order to further improve
the performance--complexity trade-off. 

\vspace{-0.2cm}

\section*{Acknowledgements}

\vspace{-0.2cm}

{\scriptsize

This work is part of a project that has received
funding from the European Union's Horizon 2020 research and innovation
programme under the Marie Sk\l{}odowska-Curie grant agreement
No.~749798. The work was also supported in part by the National
Science Foundation (NSF) under Grant No.~1609327. Any opinions,
findings, recommendations, and conclusions expressed in this material
are those of the authors and do not necessarily reflect the views of
these sponsors.

}

\vspace{-0.3cm}

\appendix


\newif\iffullbib
\fullbibfalse


\newcommand{\jlt}{J.~Lightw.~Technol.}
\newcommand{\ope}{Opt.~Exp.}
\newcommand{\tit}{IEEE Trans.~Inf.~Theory}
\newcommand{\tc}{IEEE Trans.~Comm.}
\newcommand{\ofc}{Proc.~OFC}
\newcommand{\ecoc}{Proc.~ECOC}
\newcommand{\ita}{Proc.~ITA}
\newcommand{\scc}{Proc.~SCC}


\iffullbib

\bibliographystyle{osajnl}
\bibliography{$HOME/lib/bibtex/library_mendeley}%

\else

\fi


\end{document}